\documentclass[prd,twocolumn,showpacs,amsmath]{revtex4}
\usepackage{amssymb,amsmath}
\usepackage{graphicx}
\usepackage{color}

\begin{document}
\title{Scattering by a topological defect connecting two asymptotically Minkowski spacetimes }

\author{J. P. M. Pitelli}
\email[]{joao.pitelli@ufabc.edu.br}
\affiliation{Centro de Matem\'atica, Computa\c{c}\~ao e Cogni\c{c}\~ao, UFABC, 09210-70, Santo Andr\'e, SP, Brazil}
\author{R. A. Mosna}
\email[]{mosna@ime.unicamp.br}
\affiliation{Departamento de Matem\'atica Aplicada, Universidade Estadual de Campinas, 13083-859, Campinas, SP, Brazil}

\begin{abstract}

We study the stability and the scattering properties of a spacetime with a topological defect  along a spherical bubble. This bubble connects two flat spacetimes which are asymptotically Minkowski, so that the resulting universe may be regarded as containing a wormhole. Its distinguished feature is the absence of exotic matter, i.e., its matter content respects all the energy conditions. Although this wormhole is nontraversable, waves and quantum particles can tunnel between both universes. Interestingly enough,  the wave equation alone does not uniquely determine the evolution of scalar waves on this background, and the theory of self-adjoint extensions of symmetric operators is required to find the relevant boundary conditions in this context. Here we show that, for a  particular boundary condition, this spacetime is stable and gives rise to a scattering pattern which is identical to the more usual thin-shell wormhole composed of exotic matter. Other boundary conditions of interest are also analyzed, including an unstable configuration with sharp resonances at well defined frequencies.
\end{abstract}

\maketitle

\section {Introduction}

Topological defects on cosmic scales are predicted by theories of the early universe as a result of spontaneous symmetry breaking of some unifying group \cite{vilenkin}. Cosmic strings may be generated by the breaking of a $U(1)$ symmetry, while cosmic walls may arise from the breaking of a discrete symmetry such as $\mathbb{Z}_2$ \cite{vilenkin2}. These spacetime defects are characterized by a Riemann curvature tensor which is everywhere null except on a singular submanifold. When this submanifold connects two asymptotically Minkowski spacetimes, one may interpret the corresponding solution as a wormhole \cite{letelier}.

\begin{figure}[htbp]
\begin{center}
\includegraphics[width=8.5cm]{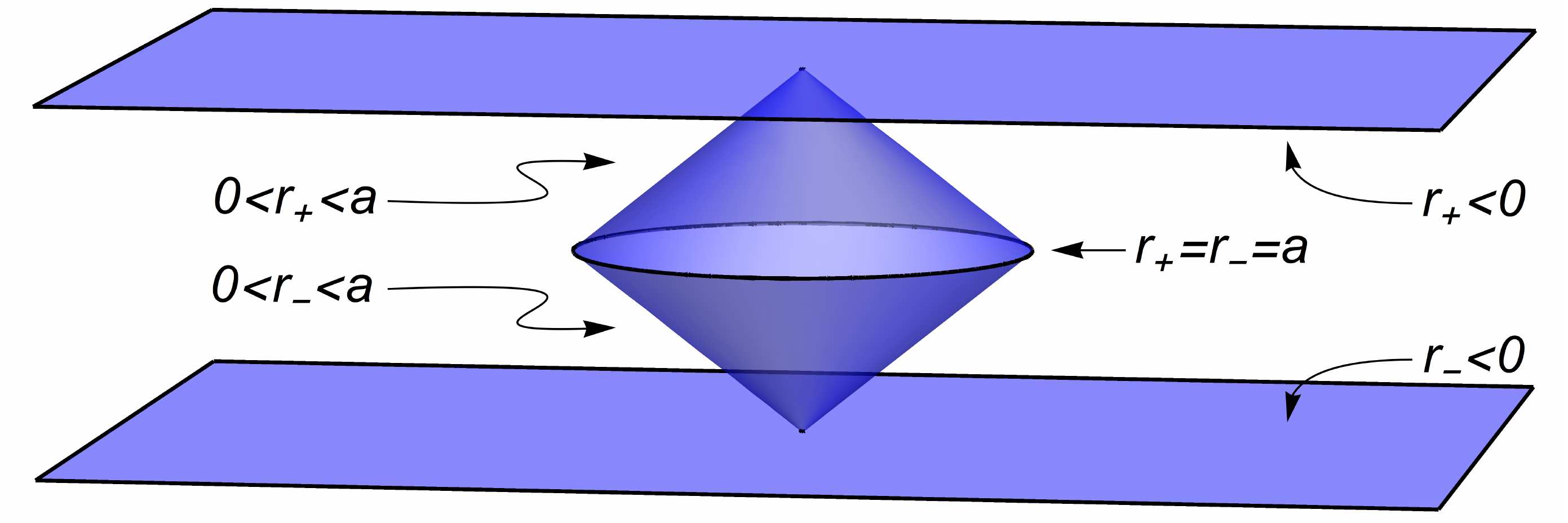}
\end{center}
\caption{Pictorial representation of the spacetime considered in this paper. We depicted the internal region as conical surfaces for illustrative purposes only; a more accurate illustration would represent it by planar discs whose boundaries are identified to each other.}
\label{fig:spacetime}
\end{figure}

In this paper, we consider a spherical topological defect constructed from two Minkowski spacetimes by (i) removing from each of them the regions described by $\Omega_{\pm}=\{r_{\pm}> a\}$, where $r_{\pm}$ represent the radial coordinate on each space and $a>0$, (ii) identifying the surfaces $\partial \Omega_{\pm}=\{r_{\pm}=a\}$ and (iii) letting the coordinates $r_{+}$ and $r_{-}$ go from $a$ to $-\infty$. Had we stopped at $r_{\pm}=0$ we would have obtained a closed baby universe. Since we extend $r_{\pm}$ up to $-\infty$ we have, instead, a baby universe connected to two asymptotically Minkowski spacetimes at two contact points ($r_{\pm}=0$), i.e., a wormhole; see Fig.~\ref{fig:spacetime}. This spacetime has been considered before, in the context of a different construction technique, by Letelier \cite{letelier}. Following the Lichnerowicz theory of distributions \cite{Lichnerowicz,Taub}, Letelier and Wang found a whole class of asymptotically Minkowski spacetimes which are locally Minkowski everywhere, except on a hypersurface with matter satisfying all the energy conditions \cite{wang}. This formalism, where flat metrics are stitched together with discontinuous derivatives which generate a thin shell of matter (a topological defect) is equivalent to the cut-and-paste procedure considered here.

With this construction, the distribution of matter at the throat $r_{\pm}=a$ respects all the energy conditions. In fact, it is composed of two families of orthogonal, spherically symmetric, Nambu strings. The price to pay for that is that this wormhole is nontraversable since an extended body cannot cross the contact points at $r_{\pm}=0$~\cite{fn1}.

However, the two ``external universes'' do communicate with each other since waves, and thus quantum particles, can pass through the contact points. The transmission pattern in this case is, remarkably, not uniquely determined by the wave equation. In fact, it turns out that the spatial portion of the wave equation in this space is not essentially self-adjoint~\cite{reed}. As a result, an infinite number of different boundary conditions are possible at the connection points, giving rise to different evolutions for the wave function.

Our aim in this paper is to explore the arbitrariness in the boundary conditions at the connection points to find candidates of stable wormholes with a topological defect as its throat. Although these wormholes must be non-traversable, we find  a particular boundary condition for which the resulting scattering pattern is the same as that of a thin-shell traversable wormhole, which requires exotic matter, found in the literature. We also show that interesting scattering resonances, at well-defined frequencies, appear for other choices of boundary conditions. Since this spacetime is everywhere flat except at the points where there is a classical singularity, the effects of the background spacetime on a quantum particle will be purely topological, with no appearance of an effective curvature potential.  In this way, its stability cannot be studied by the usual techniques employed in Refs. \cite{montelongo,visser,Lobo1,Lobo2}. Therefore, we analyze the stability of these configurations under linear perturbations of the metric by studying their quasinormal modes.

This paper is organized as follows. In Sec. II we construct the wormhole by spacetime surgery. In Sec. III we analyze the boundary conditions necessary to solve the wave equation by means of the theory of multi-interval Sturm-Liouville problems. In Sec. IV, we study the transmission and reflection coefficients of waves scattered by the wormhole, as well as its stability, and analyze their dependence on the boundary conditions found in Sec.III.  We finally discuss our results in Sec. V.

\section {Construction of the wormhole by the cut-and-paste technique}
\label{sec:construction}

Let $M_{\pm}=\mathbb{R}^3\times \mathbb{R}$ and $\Omega_{\pm}=\left\{x\in M_\pm|r_{\pm}> a\right\}$, where $r_{\pm}$ is the spatial radial coordinate in $M_{\pm}$. We remove the regions $\Omega_{\pm}$ from $M_{\pm}$ and identify the surfaces  $\partial\Omega_{\pm}=\{x|r_{\pm}=a\}\equiv \Sigma$. Finally, we extrapolate the radial coordinates $r_{\pm}$ up to $-\infty$. The resulting spacetime, $M$ is illustrated in Fig.~\ref{fig:spacetime}.

We take the usual flat metrics $ds_{\pm}^2$ on $M_{\pm}$, i.e.,
\begin{equation}
ds_{\pm}^2=-dt_{\pm}^2+dr_{\pm}^2+r_{\pm}^2d\Omega_{\pm}^2
\end{equation}
in spherical coordinates. The simple transformation of coordinates $r_\pm= \mp r +a$ then yields the following metric on $M$:
\begin{equation}
ds^2=\left\{
\begin{aligned}
&-dt^2+dr^2+(r-a)^2d\Omega^2,\,\,\,\,\, r\geq 0,\\
&-dt^2+dr^2+(r+a)^2d\Omega^2,\,\,\,\,\, r\leq 0.
\label{metric1}
\end{aligned}\right.
\end{equation}
Note that the component $g_{\theta\theta}$ of the metric is continuous but its first derivative is not. This gives rise to a thin layer of matter at $r=0$, whose content can be analyzed by using the Israel formalism of singular hypersurfaces \cite{israel}.

The induced metric at $\Sigma$, in the intrinsic coordinates $\xi^{\alpha}=(t,\theta,\varphi)$, is given by
\begin{equation}
ds^2=-dt^2+a^2d\Omega^2.
\end{equation}
The $\theta\theta$-component of the metric (\ref{metric1}),
\begin{equation}
g_{\theta\theta}=\left\{
\begin{aligned}
& (r-a)^2,\,\,\,\,\, r\geq 0,\\
& (r+a)^2,\,\,\,\,\, r\leq 0,
\end{aligned}\right.
\label{gthetatheta}
\end{equation}
is null at $r=\pm a$, so that the surfaces $r=\pm a $ have zero area, being points in space. By the Israel formalism, the vector $n_{\mu}$, normal to the singular surface, must point from $M_{-}$ to $M_{+}$, i.e.,
\begin{equation}
n^{\mu}=(0,1,0,0),
\label{normal}
\end{equation}
so that the region $M_{-}$ is given by $r<0$, while the region $M_{+}$ is given by $r>0$.

The nonzero components of the extrinsic curvature tensor at the singular surface can be written in terms of the normal derivative as
 \begin{equation}\begin{aligned}
& K_{tt}=\frac{1}{2}n^{\sigma}\frac{\partial g_{tt}}{\partial x^{\sigma}}=\frac{1}{2}\frac{\partial g_{tt}}{\partial r}=0,\\
& K_{\theta\theta}=\frac{1}{2}n^{\sigma}\frac{\partial g_{\theta\theta}}{\partial x^{\sigma}}=\frac{1}{2}\frac{\partial g_{\theta\theta}}{\partial r}=\left\{\begin{aligned}&-a\,\, \text{for}\,\, r=0^{+}\\
&+a\,\, \text{for}\,\, r=0^{-},
\end{aligned}\right.\\
& K_{\varphi\varphi}=\frac{1}{2}n^{\sigma}\frac{\partial g_{\varphi\varphi}}{\partial x^{\sigma}}=\frac{1}{2}\frac{\partial g_{\varphi\varphi}}{\partial r}=\left\{\begin{aligned}&-a\sin{\theta}\,\, \text{for}\,\, r=0^{+}\\
&+a\sin{\theta}\,\, \text{for}\,\, r=0^{-}.\end{aligned}\right.
\end{aligned}
\end{equation}
The discontinuity in the extrinsic curvature is due to the jump in the normal derivative of the metric as one crosses the defect.

Denoting $\kappa^{i}_{\phantom{i}j}=K^{i+}_{\phantom{i}j}-K^{i-}_{\phantom{i}j}$, with $K^{i\pm}_{\phantom{i}j}=K^{i}_{\phantom{i}j}$ as $r\to 0^{\pm}$,  the Lanczos equations (which come from the Einstein equations) applied to the joined surface gives the  stress-energy tensor on $\Sigma$:
 \begin{equation}
S^{i}_{\phantom{i}j}=-\left(\kappa^{i}_{\phantom{i}j}-\delta^{i}_{j}\kappa^{k}_{\phantom{k}k}\right)=\text{diag}=(2\kappa^{\theta}_{\phantom{\theta}\theta},\kappa^{\theta}_{\phantom{\theta}\theta},\kappa^{\theta}_{\phantom{\theta}\theta}).
\end{equation}
Therefore, the energy density and pressure are given by
 \begin{equation}\begin{aligned}
& \sigma=\frac{4}{a},\\
& p=-\frac{2}{a}=-\frac{\sigma}{2}.
\label{e and p}
\end{aligned}
\end{equation}

The stress energy of a classical string can be found by the Nambu-Goto action with tension $T$,
\begin{equation}
S=-T\int{d^2\xi d^4x\delta^4\left(x^\rho-X^{\rho}(\xi)\right)\sqrt{-\text{det}(h_{\alpha\beta})}},
\end{equation}
where $h_{\alpha\beta}(\xi)=\partial_{\alpha}X^{\mu}\partial_{\beta}X^{\nu}g_{\mu\nu}(X^{\mu}(\xi))$. Varying with respect to $g_{\mu\nu}$ leads to
\begin{equation}
T^{\mu\nu}=T\int{d^2\xi\delta^4\left(x^\rho-X^{\rho}(\xi)\right)h^{\alpha\beta}\partial_{\alpha}X^{\mu}\partial_{\beta}X^{\nu}}.
\end{equation}
For a meridional string  of radius $r_0$ we have $X^{\mu}=(t,r_0,\theta,0)$, thus
\begin{equation}
T^{\mu}_{\phantom{\mu}\nu}=\text{diag}\left(-T,0,-T,0\right).
\end{equation}
This yields the equation of state $p=-\sigma$. In our case we have two orthogonal directions for the strings (one direction covering the meridians and the other covering the parallels), therefore $p_{\theta}=p_{\varphi}=-\sigma/2$.

Note that this kind of matter respects all the energy conditions \cite{hawking,poisson} and, since $\rho+\sum_{i}p_i=0$, the Newtonian density is null. In this way, all the effects will be topological, with no Newtonian analogue. 

We also note that, in the above construction, the boundary $\partial\Omega_{\pm}$ of the spheres were identified in a way their interior remained an integral part of the spacetime. This procedure resulted in matter with positive energy. Had we identified those boundaries in a way that their outside remained in the spacetime we would get, instead of (\ref{gthetatheta}), 
\begin{equation}
g_{\theta\theta}=\left\{
\begin{aligned}
& (r+a)^2,\,\,\,\,\, r\geq 0,\\
& (r-a)^2,\,\,\,\,\, r\leq 0.
\end{aligned}\right.
\end{equation}
 This would result in $\sigma=-4/a$ and $p=-\sigma/2$, i.e., a layer of exotic matter.

\section{Boundary Conditions}

It follows from the metric (\ref{metric1}) that for $r>a$ and $r<-a$ we have two spaces locally isometric to Minkowski space. For $-a<r<a$ we have two balls which are connected to those universes at $r=\pm a$ and connected to each other at $r=0$. Let us define $r^{\ast}=r-a$ for $r>0$ and $r_{\ast}=-r-a$ for $r<0$. In this way the wave equation 
\begin{equation}
\left(g^{\mu\nu}\sqrt{-g}\Psi_{,\mu}\right)_{,\nu}=0
\end{equation}
reduces to its usual flat spacetime version,
\begin{equation}
\frac{d^2 R(\bar{r})}{d\bar{r}^2}+\frac{2}{\bar{r}}\frac{dR(\bar{r})}{d\bar{r}}+\left(\omega^2-\frac{l(l+1)}{\bar{r}^2}\right)R(\bar{r})=0,
\label{Sch}
\end{equation}
where we separated variables as $\Psi(t,\bar{r},\theta)=R(r)Y_{l}^{m}(\theta,\varphi)e^{-i \omega t}$ and $\bar{r}=r^{\ast}$ or $\bar{r}=r_{\ast}$.

Consider solutions of the form
\begin{equation}
u_l(r)=(\omega r)R_n(r)=\left\{
\begin{aligned}
&\eta \, e^{-i \omega r^{\ast}}+R e^{i\omega r^{\ast}},\,\,\,\,\, \text{as}\,\,\, r^{\ast}\to\infty,\\
&T e^{i \omega r_{\ast}},\,\,\,\,\, \text{as}\,\,\, r_{\ast}\to\infty,
\end{aligned}\right.
\label{first eta}
\end{equation}
where $R$ and $T$ are the reflection and transmission amplitudes, respectively. The constant $\eta$ should be set to $1$ to study scattering and to $0$ to find the quasinormal modes. For $r^{\ast}>0$ and $r_{\ast}>0$, the solutions are
\begin{equation}
R_l(r)=\left\{
\begin{aligned}
&\eta h_l^{(2)}\left(\omega r^{\ast}\right)+R h_l^{(1)}\left(\omega r^{\ast}\right),\\
&T h_l^{(1)}\left(\omega r_{\ast}\right),
\end{aligned}\right.
\end{equation}
where $h_l^{(i)}(x)$ is the spherical Hankel function of the $i$-th kind and order $l$.
Inside the balls, the general solution is a combination of spherical Hankel functions of both kinds,
\begin{equation}
R_l(r)=\left\{
\begin{aligned}
&\alpha h_l^{(1)}\left(\omega r^{\ast}\right)+\beta h_l^{(2)}\left(\omega r^{\ast}\right),\,\,\,\,\,\, r>0,\\
&\gamma h_l^{(1)}\left(\omega r_{\ast}\right)+\delta h_l^{(2)}\left(\omega r_{\ast}\right),\,\,\,\,\,\, r<0,
\end{aligned}\right.
\label{bc2}
\end{equation}
and the boundary conditions, imposed after the cut-and-paste procedure, are
\begin{widetext}
\begin{equation}
R_l(0)=\left\{\begin{aligned}
&\alpha h_l^{(1)}\left(-\omega a\right)+\beta h_l^{(2)}\left(-\omega a\right)=\gamma  h_l^{(1)}\left(-\omega a\right)+\delta h_l^{(2)}\left(-\omega a\right),\,\,\,\,\,\, \text{(continuity)}\\
&\alpha h_l^{(1)'}\left(-\omega a\right)+\beta h_l^{(2)'}\left(-\omega a\right)=-\gamma h_l^{(1)'}\left(-\omega a\right)-\delta h_l^{(2)'}\left(-\omega a\right),\,\,\,\,\,\, \text{(discontinuity of the detivatives)}
\end{aligned}\right.
\label{cont}
\end{equation}
\end{widetext}

We have three regions to consider. Those with $r>a$ and $r<-a$ correspond to two asymptotically Minkowski spacetimes. For $-a<r<a$ we have two spheres identified at the throat $\Sigma$, and the boundary condition at $\Sigma$ is enforced by Eq. (\ref{cont}). Therefore we have a three-interval Sturm-Liouville problem, with arbitrary boundary conditions to be imposed at the two connection points, $r=\pm a$. The theory of multi-interval Sturm-Liouville problem can be found in Refs. \cite{zettl,cao}, which will serve as the basis for what follows.

\subsection{Three-Interval Sturm-Liouville problem}

Let $-\infty\leq a_i\leq b_i\leq \infty$ and $J_i=(a_i,b_i)$, $i=1,2,3$. The Sturm-Liouville problem on each interval is defined by the equation
\begin{equation}
-(p_iy_i')'+q_iy_i=\lambda w_i y_i,
\label{eqgen}
\end{equation}
with weight function $w_i$ and suitable boundary conditions (in our case $p_i=r^2$, $q_i=0$ and $w_i=r^2$). The appropriate Hilbert space for this case is the direct sum $H=H_1+H_2+H_3$, where $H_i=L^2(J_i,w_i)$. Elements of $H$ are given by ${\bf f}=\{f_1,f_2,f_3\}$, $f_i\in H_i$, with inner product
\begin{equation}
({\bf f},{\bf g})=(f_1,g_1)_1+(f_2,g_2)_2+(f_3,g_3)_3.
\end{equation}
The so called Lagrange sesquilinear form is defined by
\begin{equation}\begin{aligned}
&[{\bf f},{\bf g}]=[f_1,g_1]_1(b_1)-[f_1,g_1]_1(a_1)+[f_2,g_2]_2(b_2)\\&-[f_2,g_2]_2(a_2)+[f_3,g_3]_3(b_3)-[f_3,g_3]_3(a_3),
\end{aligned}\end{equation}
where $[f,g]_i(c)=f(c) (p_i g')(c)-(p_i f')(c)g(c)$.

We are interested in the case when the outer points $a_1$ and $b_3$ are in the point limit case, i.e., when no boundary conditions are necessary there. Then, given two linear independent solutions $u(r)$ and $v(r)$ of Eq. (\ref{eqgen}) for $\lambda=0$, we define the regularized vector
\[
Y(r)=
\begin{pmatrix}
[f,u](r)\\
[f,v](r)
\end{pmatrix},
\]
where $f(r)$ is a solution of the eigenvalue problem (\ref{eqgen}).
The boundary conditions are then 	given  by
\begin{equation}
A Y(b_1)+B Y(a_2)+C Y(b_2)+D Y(a_3)=0,
\label{bc1}
\end{equation}
where  $A$, $B$, $C$ and $D$ are $4\times2$ complex matrices satisfying
\begin{itemize}
\item[(1)] $\text{rank}(A,B,C,D)=4$,
\item[(2)] $AEA^{\ast}-BEB^{\ast}+CEC^{\ast}-DED^{\ast}=0$,
\end{itemize}
with $E=\left( \begin{array}{ccc}
0 & -1  \\
1 & 0
\end{array} \right)$. In Ref. \cite{cao}, the authors showed that all the matrices satisfying conditions (1) and (2) are given, up to row and column operations, by
\begin{equation}
(A|B|C|D)=\left(\begin{array}{cccccccc}
1 & c_{11} & 0 &  c_{12} & 0 & c_{13} & 0 & c_{14}\\
0 & c_{21} & -1 &  c_{22} & 0 & c_{23} & 0 & c_{24}\\
0 & c_{31} & 0 &  c_{32} & 1 & c_{33} & 0 & c_{34}\\
0 & c_{41} & 0 &  c_{42} & 0 & c_{43} & -1 & c_{44}
\end{array}
\right),
\label{matrix1}
\end{equation}
with
\begin{equation}
\left(\begin{array}{cccccccc}
c_{11} &  c_{12}  & c_{13}  & c_{14}\\
c_{21} &  c_{22}  & c_{23} & c_{24}\\
c_{31}  &  c_{32}  & c_{33}  & c_{34}\\
c_{41}&  c_{42} &  c_{43} & c_{44}
\end{array}
\right)
\label{matrix}
\end{equation}
an arbitrary Hermitian matrix. Defining $d_1=b_1$, $d_2=a_2$, $d_3=b_2$ and  $d_{4}=a_3$, the matrix element $c_{ij}$ has the physical meaning of quantifying the ``interaction'' between the incoming wave at the point $d_i$ and the transmitted wave at the point $d_j$, $i\leq j$.

\subsection{Boundary Conditions at the Wormhole}

It is well known that Eq.~(\ref{eqgen}) is essentially self-adjoint for $l\neq 0$ so that no ambiguities arise in this case. The classical analogue of this situation is a particle with nonzero impact parameter which, as such, does not get arbitrarily close to the point $r=a$ or ($r=-a$).

The case when $l=0$ is more interesting since a particle with zero angular momentum will travel directly towards a connection point. A classical particle arriving at such a point ($r=a$ or $r=-a$) needs new initial conditions (Does it reflect back to its universe? Does it reach the throat?). This ambiguity appears at the quantum level as well. Two linear independent square-integrable  solutions exist near $r=\pm a$ and thus extra boundary conditions will be necessary at these points \cite{fn3}.

Given the eigenvalue problem for $l=0$,
\begin{equation}
-\left(r^2 R'(r)\right)'=\lambda r^2 R(r),
\end{equation}
two linear independent solutions for $\lambda=0$ are given by $u(r)=1$ and $v(r)=\frac{1}{r}$ . Then
\[
Y(r)=\left(\begin{array}{cc}
&[R,u](r)\\&[R,v](r)\end{array}\right)=\left(\begin{array}{cc}
&-(r^2 R'(r))\\&-R(r)-rR'(r)\end{array}\right).
\]

In the present case our three interval problem is defined by the intervals $(-\infty,- a)$, $(-a,a)$ and $(a,\infty)$, and the outer points $\pm\infty$ are in the limit point case. If we define $a^{\pm}=a\pm0^+$ and $-a^{\pm}=-a\pm 0^+$ we have (by using Eq. (\ref{first eta})):
\begin{widetext}
\begin{equation}\left\{\begin{aligned}
&R(r)=\eta\frac{e^{-i \omega r^{\ast}}}{r^{\ast}}+R \frac{e^{i \omega r^{\ast}}}{r^{\ast}}, \,\,\, r>a\,\,\,\Rightarrow Y(a^{+})=\left(\begin{array}{cc}
&\eta+R\\&i \eta \omega - i\omega R\end{array}\right),\\
&R(r)=\alpha \frac{e^{-i \omega r^{\ast}}}{r^{\ast}}+\beta \frac{e^{i \omega r^{\ast}}}{r^{\ast}}, \,\,\, 0<r<a\,\,\, \Rightarrow Y(a^{-})=\left(\begin{array}{cc}
&\alpha+\beta\\&i \omega \alpha - i\omega \beta\end{array}\right),\\
&R(r)=\gamma \frac{e^{-i \omega r_{\ast}}}{r_{\ast}}+\delta \frac{e^{i \omega r_{\ast}}}{r_{\ast}}, \,\,\, -a<r<0\,\,\, \Rightarrow Y(-a^{+})=\left(\begin{array}{cc}
&\gamma+\delta\\&-i \gamma \omega +i\omega \delta\end{array}\right),\\
&R(r)=T \frac{e^{i \omega r_{\ast}}}{r_{\ast}}, \,\,\, r<-a\,\,\, \Rightarrow Y(-a^{-})=\left(\begin{array}{cc}
&T\\&i \omega T\end{array}\right),
\end{aligned}\right.
\label{wave form}
\end{equation}
\end{widetext}
where the orientation of the radial coordinate has been taken into account and $\eta$ was defined right after Eq.~(\ref{first eta}).

As a result, the boundary conditions are given by Eq.(\ref{cont}) together with
\begin{equation}
A Y(a^{+})+B Y(a^{-})+C Y(-a^{+})+D Y(-a^{-})=0.
\label{bc3}
\end{equation}

\section{Scattering and Stability}
\label{seciii}

In this section we study (i) the scattering of waves by the wormhole constructed in Section II and (ii) the stability of the corresponding spacetime under linear perturbations of the metric. As discussed above, for $l\neq 0$ the wave equation is already self-adjoint so that, in this case, no boundary conditions are necessary at $r=\pm a$. This means that partial waves with nonzero angular momentum are completely unaware of the wormhole existence. We therefore restrict ourselves to spherical waves and set $l=0$ in what follows \cite{fn_qm}.

The scattering patterns arising from this problem fall essentially into two broad classes. The first of them corresponds to physically acceptable boundary conditions wherein all the interactions are local, i.e., an incoming wave at $r=a^{+}$ ($r=-a^{+}$) interacts only with the transmitted wave at $r=a^{-}$ ($r=-a^{-}$). The second class comprises boundary conditions which are mathematically possible but physically unattainable due to violation of causality (as, for instance, a direct interaction between  $r=a^{+}$ and $r=-a^{-}$); this case is analyzed, for the sake of completeness, in the Appendix. The remaining of this section deals with the first class.

For local interactions, (\ref{matrix}) assumes the form of a $2\times2$ block diagonal matrix. The physics here can be understood by analyzing two prototypical cases. In the first case, the boundary conditions guarantee continuity of the wave function and its derivative at $r=\pm a$. The classical analogue of this situation is a classical particle passing directly through $r=\pm a$ without ``changing direction''. In the second case, the boundary conditions mix the values of the wave function and its derivative at $a^+$ and $a^-$ (also at  $-a^{+}$ and $-a^{-}$). Classically, this corresponds to a particle following a radial trajectory, which pass through $r=\pm a $ possibly ``changing its direction'', but still following a radial line. The nonuniqueness of boundary conditions in the quantum domain allows both possibilities.

We also investigate the stability of these configurations by finding the characteristic quasinormal modes of the wormhole, which also depend on the boundary conditions. It is well known that quasinormal modes appear when the metric is perturbed and the outgoing-wave boundary condition at both universes is imposed \cite{Kokkotas}. For a wave function with time dependence given by $e^{-i \omega t}$, an $\omega$ with negative imaginary part indicates that the perturbation is damped, otherwise the perturbation grows and the mode is dynamically unstable.

The calculations that follow are based on Eq. (\ref{wave form}). For the scattering analysis, we impose scattering boundary conditions at infinity so that $\eta=1$. We then get four equations from Eq. (\ref{bc1}) plus two equations from Eq. (\ref{cont}) with six unknowns to be determined ($R$, $\alpha$, $\beta$, $\gamma$, $\delta$ and $T$). On the other hand, the quasinormal modes require outgoing-wave boundary conditions on both external universes. This corresponds to $\eta=0$ so that we have more equations than unknowns. The quasinormal modes are then calculated by requiring that the linear system resulting from the boundary conditions has a nontrivial solution, i.e., that its determinant is null.
The calculations that follow were done with the help of the software Mathematica \cite{Mathematica}.

\subsection{First Case}
\label{subsec:first}

Here we impose continuity of the wave function and its derivatives at $r=\pm a$.  The matrix $(A,B,C,D)$ associated with this boundary conditions is given by
\begin{equation}
\left(\begin{array}{cccccccc}
1 &  0  & -1 & 0 &0  & 0  & 0 &0\\
0 &  -1  & 0 & 1 &0   & 0  & 0 &0\\
0  &  0  & 0  & 0&1   &  0 &-1 &0\\
0&  0    &  0 & 0&0    &  -1 &0&1
\end{array}
\right).
\end{equation}
We note that this matrix is constructed by taking
\begin{equation}
\left(\begin{array}{cccccccc}
-1 &  0  & 0 & 0\\
0 & 1 & 0 & 0\\
0  & 0   & -1  & 0\\
0&  0 &  0 & 1
\end{array}
\right)
\end{equation}
as (\ref{matrix})  and then performing column operations on (\ref{matrix1}). Taking $\eta=1$ in Eq. (\ref{wave form}) (scattering), we get boundary conditions given by
\begin{equation}\begin{aligned}
&A Y(a_1)+B Y(a_2)+C Y(a_3)+D Y(a_4)\\&=\left(\begin{array}{c}
1+R-\alpha-\beta\\
i(-1+R+\alpha-\beta)\omega\\
-T+\gamma+\delta\\
i(T+\gamma-\delta)\omega
\end{array}\right)=
\left(\begin{array}{c}
0\\
0\\
0\\
0
\end{array}\right).\end{aligned}
\end{equation}

\begin{figure}[htbp]
\begin{center}
\includegraphics[width=7.7cm]{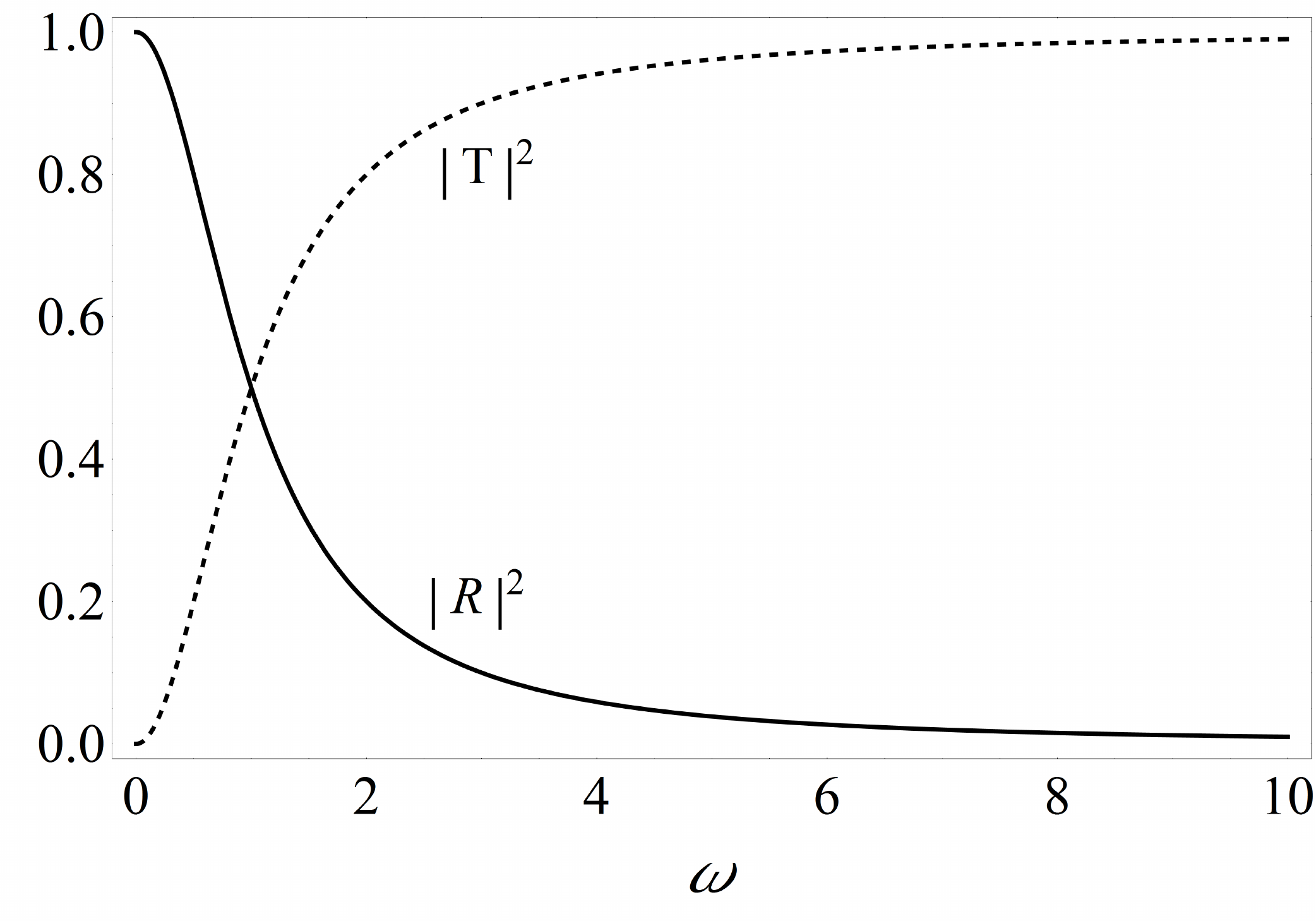}
\end{center}
\caption{Transmission (dashed line) and reflection (solid line) probabilities as functions of the frequency $\omega$ for the case considered in Section \ref{subsec:first}. Units were chosen such that the radius of the throat, $a$, is $1$.}
\label{fig1}
\end{figure}

A direct calculation shows that this (together with the boundary condition at the throat, Eq. (\ref{cont})) yields the following reflection and transmission amplitudes:
\begin{equation}\begin{aligned}
&R=\frac{e^{-2 i \omega }}{i \omega -1 },\\
&T=\frac{i \omega e^{-2 i \omega } }{i \omega - 1}.
\end{aligned}
\end{equation}
The corresponding transmission and reflection probabilities are show in Fig. \ref{fig1}.

This scattering pattern is the same as that of a one-dimensional particle with energy $\omega$ subjected to a delta potential with strength $1/a$. This can be understood by noticing that a spherical wave passes through $r=\pm a$ with no deflection except at the throat $\Sigma$ where the sign of the radial derivative changes abruptly. An identical pattern is seen in a simplified version of standard traversable wormholes wherein the throat is sustained by strings with negative tension \cite{Khabibullin,Khusnutdinov}. In those works the authors start with two identical copies of Minkowski spacetime, excise from each of them a spherical region, and then identify the resulting boundaries. With that construction, the singular curvature at the throat also gives rise to an effective delta potential. One should note, however, that the scattering is not restricted to waves of zero angular momentum in that model.

To find the quasinormal modes for our case we take $\eta=0$ in Eq. (\ref{wave form}). The determinant $D$ of the resulting homogeneous linear system is given by
\begin{equation}
D=8 i e^{2 i \omega } \omega ^2 (\omega +i),
\end{equation}
with roots $\omega=0$ and $\omega=-i$. Since the (single) dynamical mode has negative imaginary part, the spacetime in this case is stable under linear perturbations.

\subsection{Second Case}
\label{subsec:second}

Here we consider boundary conditions which result in nontrivial scattering of the incident wave at $r=\pm a$.  A typical representative for his case is a matrix $(A,B,C,D)$ given by
\begin{equation}
\left(\begin{array}{cccccccc}
1 &  0  &  0 & 1 & 0  & 0  & 0 &0\\
0 &  1  & -1 & 0 & 0  & 0  & 0 &0\\
0 &  0  &  0 & 0 & 1  & 0  & 0 &1\\
0 &  0  &  0 & 0 & 0  & 1  &-1 &0
\end{array}
\right),
\end{equation}
which leads to (by Eq. (\ref{wave form}))
\begin{equation}\begin{aligned}
&A Y(a_1)+B Y(a_2)+C Y(a_3)+D Y(a_4)=\\&\left(\begin{array}{c}
1+R+i \omega(\alpha-\beta)\\
-\alpha-\beta-i\omega(-1+R)\\
\gamma+\delta+i\omega T\\
-T-i \omega(\gamma-\delta)
\end{array}\right)=
\left(\begin{array}{c}
0\\
0\\
0\\
0
\end{array}\right).\end{aligned}
\end{equation}
Note that the wave function and its derivatives are mixed at $a^{+}$ and $a^{-}$ (also at $-a^{+}$ and $-a^{-}$). The reflection and transmission amplitudes are then given by
\begin{equation}\begin{aligned}
&R=\tfrac{2 e^{2 i \omega } \left(\omega ^4+\left(\omega ^4-1\right) \omega  \sin (2 \omega )+\left(\omega ^4-1\right) \cos (2 \omega )+1\right)}{2 e^{2 i \omega } \left(\omega ^4-1\right)+(1+i \omega ) \left(\omega ^2-1\right)^2+e^{4 i \omega } (\omega -i)^2 (-1+i \omega )^3},\\
&T=-\tfrac{4 e^{2 i \omega } \omega ^3}{2 i e^{2 i \omega } \left(\omega ^4-1\right)-(\omega -i) \left(\omega ^2-1\right)^2+e^{4 i \omega } (\omega -i)^2 (\omega +i)^3}.
\end{aligned}
\end{equation}

\begin{figure}[htbp]
\begin{center}
\includegraphics[width=7.7cm]{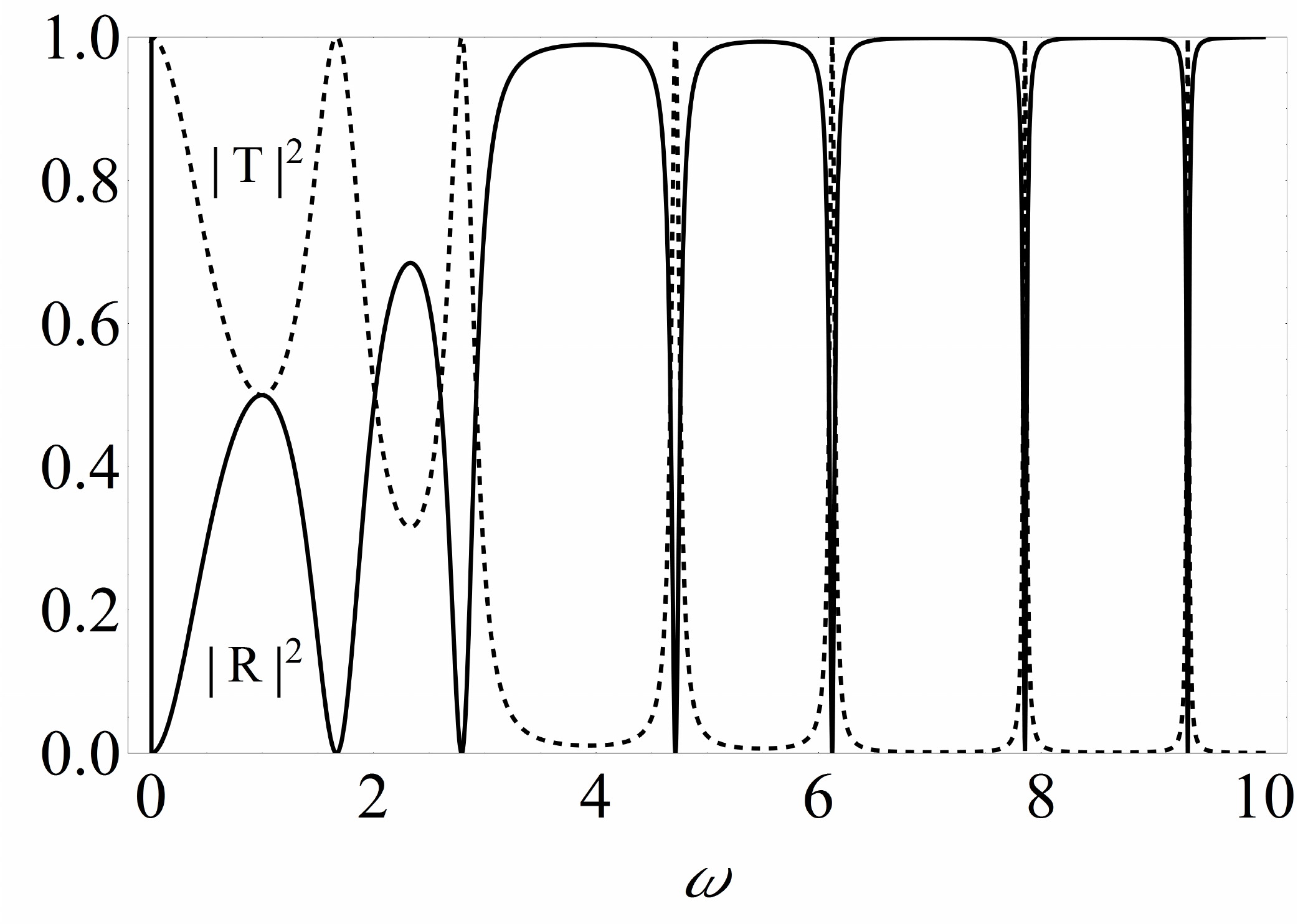}
\end{center}
\caption{Transmission (dashed line) and reflection (solid line) probabilities as functions of the frequency $\omega$ for the case considered in Section \ref{subsec:second}. Units were chosen such that the radius of the throat, $a$, is $1$.}
\label{fig3}
\end{figure}

It is interesting to note the emergence of resonances in this case, with peaks of transmission at well defined frequencies; see Fig. \ref{fig3}. This happens when the derivative of the wave function vanishes at the throat. By means of Eq. (\ref{cont}), no effective potential appears at the throat so that only the boundary conditions at the connection points contribute to the scattering. Clearly, as $\omega\to \infty$, $T\to 1$ at the resonances. Since $h_0^{(1)}(\omega r)$ and $h_0^{(2)}(\omega r)$ behave as linear combinations of $\sin\left(\omega r-\pi/4\right)/\sqrt{\omega r}$ and $\cos\left(\omega r-\pi/4\right)/\sqrt{\omega r}$ for $\omega\gg1$, these resonances will be periodic in this limit. Notice that the size of the wormhole could then be inferred from the scattering data.

\begin{figure}[htbp]
\begin{center}
\includegraphics[width=7.7cm]{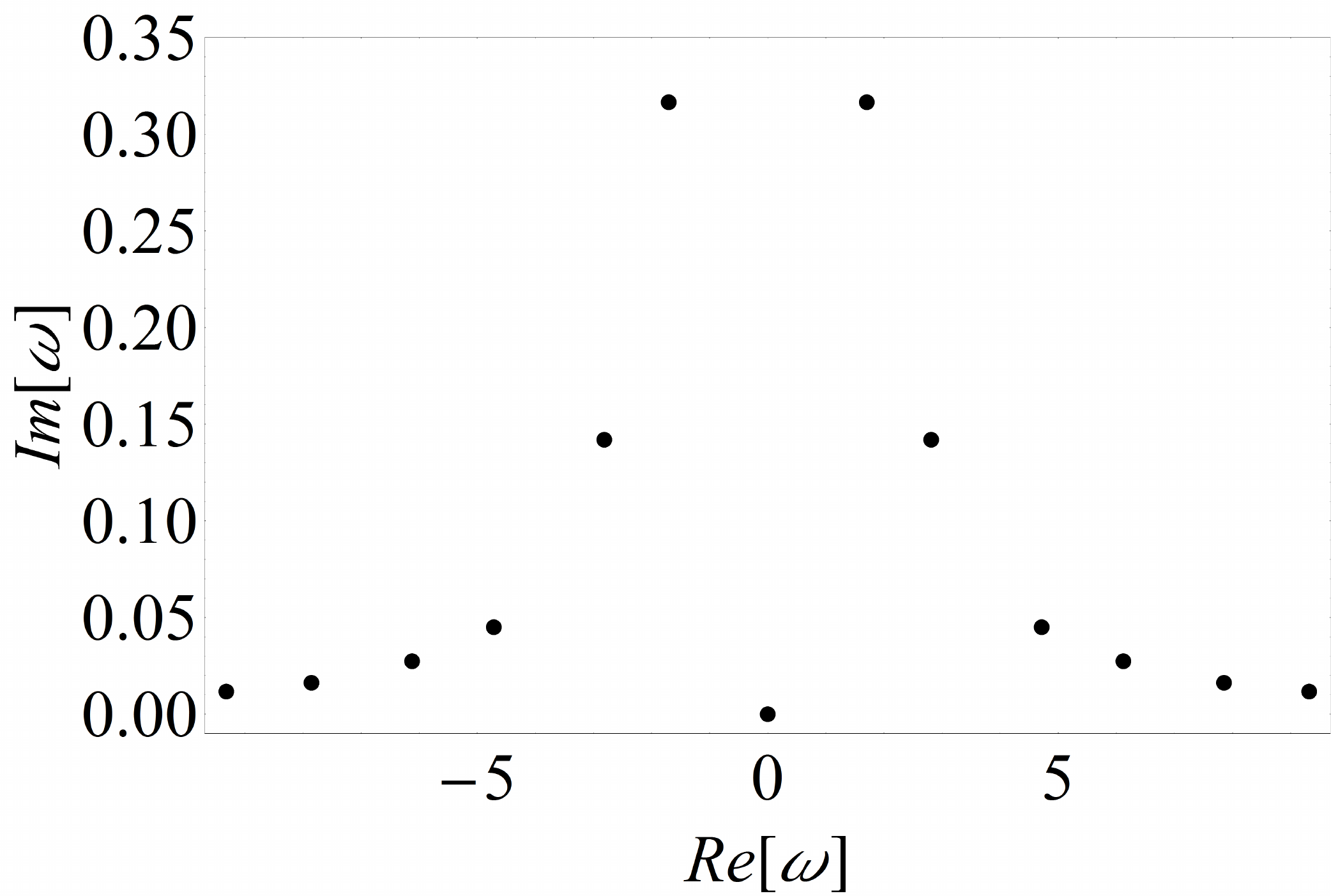}
\end{center}
\caption{Quasinormal modes for the case considered in Section \ref{subsec:second}. Units were chosen such that the radius of the throat, $a$, is $1$.}
\label{fig5}
\end{figure}

However, with such boundary conditions the wormhole is unstable. In fact, the determinant $D$ of the homogeneous linear system composed of Eqs. (\ref{cont}) and (\ref{bc3})  is given by
\begin{equation}
\begin{aligned}
D&=4 \left[\omega ^4+\left(\omega ^4+2 i \omega +1\right) \omega  \sin (2 \omega )\right.+\\ & \qquad \left. {}+\left(1+(\omega -2 i) \omega ^3\right) \cos (2 \omega )-1\right].
\end{aligned}
\end{equation}
The numerically determined roots of this equation are shown in Fig.~\ref{fig5}, from which we see that imaginary part of the quasinormal modes is indeed positive.

\section{Conclusions}

We studied the problem of wave propagation on a wormhole supported by two orthogonal families of Nambu strings. As such, 
the wormhole considered here has a throat sustained by matter which respects all the energy conditions. After finding all the boundary conditions that guarantee unitary evolution for the wave equation on this background, we studied how they affect the scattering pattern and the stability of this spacetime under linear perturbations of the metric. Two particular examples of such boundary conditions were analyzed in detail: one for which the scattering pattern of spherical waves is virtually indistinguishable from that of simplified versions of the more usual wormholes sustained by exotic matter, and another with resonances and a highly oscillatory behavior. For these examples, we found that the spacetime is stable and unstable, respectively.

Other examples of wormholes sustained by Nambu strings with different symmetries can also be probed by waves and analyzed by the methods presented here. Some examples are cylindrical and parabolic plates, as well as parabolic and toroidal shells (see Ref. \cite{letelier}). In the particular case of the cylindrical plate, the throat is sustained by parallel fibers in the direction of the $z$-axis (a tube of cosmic strings). Perhaps surprisingly, this spacetime is also asymptotically flat (and not asymptotically conical). We hope to address this case in a future publication.

\acknowledgements
We acknowledge stimulating discussions with M. Richartz and A. Saa and the comments of an anonymous referee. This work was partially supported by FAPESP grant 2013/09357-9.

\appendix
\section{}
\label{app1}

Here we study, for completeness' sake, a typical example of the second class of boundary conditions discussed in Sec. \ref{seciii}. In this case the wave function at $a^{+}$ interacts directly with the wave function at $-a^{-}$. This can be accomplished by letting the elements $c_{14}$ and $c_{41}$ in (\ref{matrix}) be nonzero.

The matrix $(A,B,C,D)$ related to this boundary conditions is given by
\begin{equation}
\left(\begin{array}{cccccccc}
1 &  0  & 0 & 0 &0  & 0  & 0 &1\\
0 &  0  & -1 & 0 &0   & 0  & 0 &0\\
0  &  0  & 0  & 0&1   &  0 &0 &0\\
0&  1    &  0 & 0&0    &  0 &-1&0
\end{array}
\right),
\end{equation}
which corresponds to the following Hermitian matrix (as in Eq. (\ref{matrix})):
\begin{equation}
\left(\begin{array}{cccccccc}
0 &  0  & 0 & 1\\
0 & 0 & 0 & 0\\
0  & 0   & 0  & 0\\
1&  0 &  0 & 0
\end{array}
\right).
\end{equation}
The resulting boundary condition is then given by
\begin{equation}\begin{aligned}
&A Y(a_1)+B Y(a_2)+C Y(a_3)+D Y(a_4)\\&=\left(\begin{array}{c}
1+R+i \omega T\\
-\alpha-\beta\\
\gamma+\delta\\
-T-i \omega(-1+R)
\end{array}\right)=
\left(\begin{array}{c}
0\\
0\\
0\\
0
\end{array}\right).\end{aligned}
\end{equation}

\begin{figure}[htbp]
\begin{center}
\includegraphics[width=7.7cm]{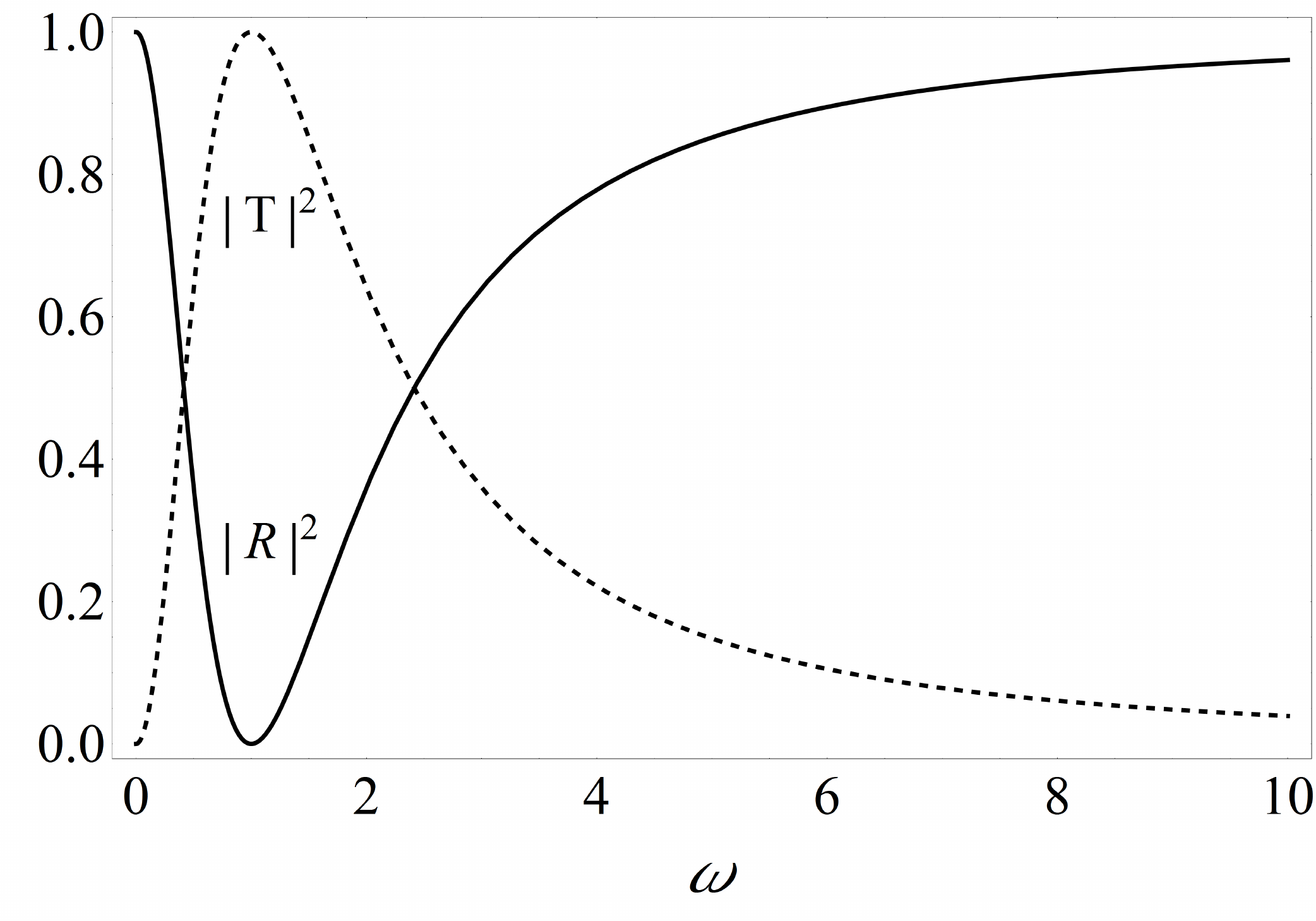}
\end{center}
\caption{Transmission (dashed line) and reflection (solid line) probabilities as functions of the frequency $\omega$ for the case considered in the Appendix. Units were chosen such that the radius of the throat, $a$, is $1$.}
\label{fig6}
\end{figure}

The reflection and transmission amplitudes now become
\begin{equation}\begin{aligned}
&R=\frac{\omega^2-1}{\omega^2+1},\\
&T=\frac{2 i \omega }{\omega ^2+1}.
\end{aligned}
\end{equation}
The corresponding transmission and reflection probabilities are shown in Fig. \ref{fig6}.

We note that what happens in this case is a direct interaction between the incident wave  at $a^{+}$ and the transmitted wave at $-a^{-}$. The wave function is thus zero in the region defined by $-a<r<a$. This can be interpreted as an instant transmission between two distant points, which violates causality. One way to give physical meaning to this kind of nonlocal boundary condition is to consider a symmetric configuration with waves coming from both $r=-\infty$ and $r=+\infty$. The resulting standing wave pattern, although not a scattering configuration, would then be physically acceptable.


\begin{thebibliography}{99}
\bibitem{fn1} After all, traversable wormholes are known to require exotic matter \cite{morris}, i.e., matter that violates the energy conditions (other, more stringent conditions are also necessary to guarantee stability \cite{montelongo}).
\bibitem{fn3} Since  boundary conditions are necessary at $r=\pm a$, but not at $r^{\ast}=\infty$ and $r_{\ast}=\infty$,  the first and third intervals have deficiency index $d=1$, while the second has deficiency index $d=2$. Therefore, the deficiency index on the whole spacetime is given by $d=1+2+1=4$.
\bibitem{fn_qm} This analysis deals with the scalar case only. One could speculate what would happen in a true quantum mechanical setting, where the contact points are in principle smeared out and particles with large angular momenta and large energies could cross the region. These considerations are, however, beyond our semiclassical analysis.
\end{thebibliography}
\end{document}